\documentclass[a4paper,11pt]{article}
\usepackage{pos}
\usepackage{amsmath}

\title{Measurement of the W boson mass at LHCb}

\author*[a, 1]{Ross Hunter}

\affiliation[a]{University of Warwick, United Kingdom}

\note{On behalf of the LHCb collaboration}

\emailAdd{ross.john.hunter@cern.ch}

\abstract{Constraints on new physics in the electroweak sector are limited by the precision of direct measurements of the \PW boson mass ($m_W$). A new measurement is hereby reported, using proton-proton collision data recorded by the LHCb experiment in 2016 at $\sqrt{s}=13\tev$, corresponding to roughly 1.7 $\textrm{fb}^{-1}$ of integrated luminosity. From a simultaneous fit of the muon $q/\pt$ distribution from $W \to \mu\nu$ decays and the $\phi^{\ast}$ distribution from $Z \to \mu\mu$ decays, $m_W$ is measured to be 

\begin{equation*}
    m_W = \DataMWValue \pm \DataMWStat_{\textrm{stat}} \pm \DataMWExp_{\textrm{exp}} \pm \DataMWTh_{\textrm{theory}} \pm \DataMWPDF_{\textrm{PDF}} \,\mev,
\end{equation*}

where the uncertainties are due to statistical, experimental systematic, theoretical and parton distribution function sources respectively. This is an average of
results based on three recent global parton distribution function sets, and is compatible with previous measurements as well as the prediction from the global electroweak fit. This measurement is a pathfinder for a full Run-2 (2016-2018) measurement from LHCb, which is expected to be competitive with current world-leading measurements, and to make a substantial contribution to an LHC-wide average due to the complementary acceptance of \lhcb with respect to ATLAS and CMS.}

\FullConference{%
  *** The European Physical Society Conference on High Energy Physics (EPS-HEP2021), ***\\
  *** 26-30 July 2021 ***\\
  *** Online conference, jointly organized by Universität Hamburg and the research center DESY ***
}

\usepackage{xspace} 
\usepackage{upgreek}

\def\lhcb   {\mbox{LHCb}\xspace}
\def\atlas  {\mbox{ATLAS}\xspace}
\def\cms    {\mbox{CMS}\xspace}

\def\cdf    {\mbox{CDF}\xspace}
\def\dzero  {\mbox{D0}\xspace}
\def\aleph  {\mbox{ALEPH}\xspace}
\def\delphi {\mbox{DELPHI}\xspace}
\def\opal   {\mbox{OPAL}\xspace}
\def\lthree {\mbox{L3}\xspace}

\def\MagUp {\mbox{\em Mag\kern -0.05em Up}\xspace}

{

 \def\Ppsi        {\ensuremath{\uppsi}\xspace}

 \def\PDelta      {\ensuremath{\Delta}\xspace}                 
 \def\PXi         {\ensuremath{\Xi}\xspace}                 
 \def\PLambda     {\ensuremath{\Lambda}\xspace}                 
 \def\PSigma      {\ensuremath{\Sigma}\xspace}                 
 \def\POmega      {\ensuremath{\Omega}\xspace}                 
 \def\PUpsilon    {\ensuremath{\Upsilon}\xspace}

 \def\PB      {\ensuremath{\mathrm{B}}\xspace}                 
                  
 \def\PD      {\ensuremath{\mathrm{D}}\xspace}

 \def\PJ      {\ensuremath{\mathrm{J}}\xspace}                 
 \def\PK      {\ensuremath{\mathrm{K}}\xspace}

 \def\PW      {\ensuremath{\mathrm{W}}\xspace}

 \def\PZ      {\ensuremath{\mathrm{Z}}\xspace}

 \def\Pi      {\ensuremath{\mathrm{i}}\xspace}

 \def\Ps      {\ensuremath{\mathrm{s}}\xspace}

 \def\thebaroffset{0.0em}
}
{

 \def\Ppsi        {\ensuremath{\psi}\xspace}                 
                  
 \mathchardef\PDelta="7101
 \mathchardef\PXi="7104
 \mathchardef\PLambda="7103
 \mathchardef\PSigma="7106
 \mathchardef\POmega="710A
 \mathchardef\PUpsilon="7107
                  
 \def\PB      {\ensuremath{B}\xspace}                 
                  
 \def\PD      {\ensuremath{D}\xspace}

 \def\PJ      {\ensuremath{J}\xspace}                 
 \def\PK      {\ensuremath{K}\xspace}

 \def\PW      {\ensuremath{W}\xspace}

 \def\PZ      {\ensuremath{Z}\xspace}

 \def\Pi      {\ensuremath{i}\xspace}

 \def\Ps      {\ensuremath{s}\xspace}

 \def\thebaroffset{0.18em}
}
\newcommand{\offsetoverline}[2][\thebaroffset]{\kern #1\overline{\kern -#1 #2}}%

\makeatletter
\ifcase \@ptsize \relax
  \newcommand{\miniscule}{\@setfontsize\miniscule{4}{5}}
\or
  \newcommand{\miniscule}{\@setfontsize\miniscule{5}{6}}
\or
  \newcommand{\miniscule}{\@setfontsize\miniscule{5}{6}}
\fi
\makeatother

\DeclareRobustCommand{\optbar}[1]{\shortstack{{\miniscule (\rule[.5ex]{1.25em}{.18mm})}
  \\ [-.7ex] $#1$}}

\def\squark    {{\ensuremath{\Ps}}\xspace}

\def\KorKbar {\kern \thebaroffset\optbar{\kern -\thebaroffset \PK}{}\xspace}

\def\D       {{\ensuremath{\PD}}\xspace}

\def\DorDbar {\kern \thebaroffset\optbar{\kern -\thebaroffset \PD}\xspace}

\def\Dp      {{\ensuremath{\D^+}}\xspace}
\def\Dm      {{\ensuremath{\D^-}}\xspace}

\def\DpDm    {\ensuremath{\Dp {\kern -0.16em \Dm}}\xspace}

\def\B       {{\ensuremath{\PB}}\xspace}

\def\BorBbar {\kern \thebaroffset\optbar{\kern -\thebaroffset \PB}\xspace}

\def\Bd      {{\ensuremath{\B^0}}\xspace}

\def\BdorBdbar {\kern \thebaroffset\optbar{\kern -\thebaroffset \Bd}\xspace}

\def\Bs      {{\ensuremath{\B^0_\squark}}\xspace}

\def\BsorBsbar {\kern \thebaroffset\optbar{\kern -\thebaroffset \Bs}\xspace}

\def\jpsi     {{\ensuremath{{\PJ\mskip -3mu/\mskip -2mu\Ppsi}}}\xspace}

\def\Upsilonres  {{\ensuremath{\PUpsilon}}\xspace}
\def\Y#1S{\ensuremath{\PUpsilon{(#1S)}}\xspace}

\def\LorLbar     {\kern \thebaroffset\optbar{\kern -\thebaroffset \PLambda}\xspace}

\def\to                 {\ensuremath{\rightarrow}\xspace}

\def\AT#1     {\ensuremath{A_{\mathrm{T}}^{#1}}\xspace}           

\def\C#1      {\ensuremath{\mathcal{C}_{#1}}\xspace}                       
\def\Cp#1     {\ensuremath{\mathcal{C}_{#1}^{'}}\xspace}                    
\def\Ceff#1   {\ensuremath{\mathcal{C}_{#1}^{\mathrm{(eff)}}}\xspace}        
\def\Cpeff#1  {\ensuremath{\mathcal{C}_{#1}^{'\mathrm{(eff)}}}\xspace}       
\def\Ope#1    {\ensuremath{\mathcal{O}_{#1}}\xspace}                       
\def\Opep#1   {\ensuremath{\mathcal{O}_{#1}^{'}}\xspace}                    

\newcommand{\aunit}[1]{\ensuremath{\text{\,#1}}}       

\newcommand{\tev}{\aunit{Te\kern -0.1em V}\xspace}
\newcommand{\gev}{\aunit{Ge\kern -0.1em V}\xspace}
\newcommand{\mev}{\aunit{Me\kern -0.1em V}\xspace}
\newcommand{\kev}{\aunit{ke\kern -0.1em V}\xspace}
\newcommand{\ev}{\aunit{e\kern -0.1em V}\xspace}
 
\newcommand{\mevc}{\ensuremath{\aunit{Me\kern -0.1em V\!/}c}\xspace}
\newcommand{\gevc}{\ensuremath{\aunit{Ge\kern -0.1em V\!/}c}\xspace}
\newcommand{\mevcc}{\ensuremath{\aunit{Me\kern -0.1em V\!/}c^2}\xspace}
\newcommand{\gevcc}{\ensuremath{\aunit{Ge\kern -0.1em V\!/}c^2}\xspace}

\def\gsim{{~\raise.15em\hbox{$>$}\kern-.85em
          \lower.35em\hbox{$\sim$}~}\xspace}
\def\lsim{{~\raise.15em\hbox{$<$}\kern-.85em
          \lower.35em\hbox{$\sim$}~}\xspace}

\def\pt         {\ensuremath{p_{\mathrm{T}}}\xspace}

\def\pythia     {\mbox{\textsc{Pythia}}\xspace}

\def\powhegbox     {\mbox{\textsc{POWHEGBoxV2}}\xspace}

\def\tell1  {TELL1\xspace}
\def\ukl1   {UKL1\xspace}

\def\DefaultFigWidth{0.49\textwidth}
\def\IKT    {{\ensuremath{k_{\rm T}^{\rm intr}}}\xspace}
\def\PW      {\ensuremath{W}\xspace}
\def\PZ      {\ensuremath{Z}\xspace}
\def\PUpsilon    {\ensuremath{\Upsilon}\xspace} 
\def\DYTurbo     {\mbox{\textsc{DYTurbo}}\xspace}
\def\Ppsi        {\ensuremath{\uppsi}\xspace}  
\def\PJ      {\ensuremath{\mathrm{J}}\xspace} 
\def\DataMWValue {80354}

\def\DataMWStat {23}
\def\DataMWExp {10}
\def\DataMWTh {17}
\def\DataMWPDF {9}

\begin{document}
\maketitle

\section{Introduction}

\noindent
In the Standard Model (SM), the mass of the \PW boson ($m_W$) can be predicted in a global fit to the parameters of the electroweak (EW) sector. New physics in this sector can therefore be constrained/inferred by comparing with direct measurements of $m_W$. At present, the global EW fit predicts $m_W$ with a 7\mev uncertainty~\cite{Gfitter} -- almost half the uncertainty of the 2020 PDG average of direct measurements (12\mev)~\cite{PDG2020}. This provides strong motivation for new, high-precision direct measurements of $m_W$.

Previously, $m_W$ has been measured to a precision of 33\mev at LEP~\cite{LEP:ew:2013} and 16\mev at the Tevatron~\cite{Tevatron}. The sole LHC measurement before now was performed by \atlas, achieving a 19\mev uncertainty~\cite{atlas}. Despite using only a small subset of the \atlas data collected to date, this measurement was already limited by the modelling of \PW boson production, in particular by the uncertainties in the proton's parton distribution functions (PDFs). Ref.~\cite{Vesterinen} showed that, if a measurement from \lhcb were averaged with one from \atlas or \cms, the PDF-related uncertainty would partially cancel. This is due to the complementary pseudorapidity ($\eta$) acceptance of \lhcb: it is a single-arm spectrometer, fully instrumented in the "forward" region $2 < \eta < 5$~\cite{LHCb}. Furthermore, it was estimated that with the full Run-2 (2016-2018) \lhcb data, a statistical precision of 10\mev would be achievable~\cite{Vesterinen}. However, at this time, theoretical uncertainties in the \PW boson production model would be a limiting factor in achieving a competitive overall precision. Our goal was therefore to measure $m_W$ using the 2016 data only, to pave the way for further collaboration and effort towards the ultimate Run-2 precision measurement.

\section{Analysis Strategy}

In $W \to \mu\nu$ decays, the muon transverse momentum (\pt) has a characteristic ``Jacobian edge'' at $\sim m_W/2$. This allows $m_W$ extraction with a template fit to the muon \pt distribution: simulated templates are prepared using different values of $m_W$, and the best-fitting template corresponds to the best-fitting $m_W$. The challenge of such a measurement is the accurate simulation of these templates, and controlling the associated uncertainties. The leading theoretical contributions to this are in the modelling of \PW boson production and decay, which is described by a 5D differential cross section, and further factorised into an unpolarised cross section and an angular distribution. Here, the former is parametrised by the boson transverse momentum ($\pt^V$), rapidity ($y$) and mass ($M$), while the latter is written in terms of two decay angles and eight \textit{angular coefficients} ($A_0 - A_7$). Here, the angular coefficients (which are ratios of helicity cross sections) are calculated at $\mathcal{O}(\alpha_s^2)$ using the \DYTurbo program~\cite{DYTurbo2020}. Propagating the uncertainties from these calculations -- particularly of $A_3$ -- to $m_W$ initially yielded a dominating $\mathcal{O}(30)\mev$ uncertainty, which was reduced by introducing a floating $A_3$ scale factor in the fit.

Our central model of the unpolarised cross section is provided by \powhegbox~\cite{POWHEGewW,POWHEGewZ}, interfaced to \pythia8~\cite{Pythia8Main} for simulation of the parton shower. Previous $m_W$ measurements have relied on tuning event generators to the $\pt^Z$ distribution in $Z \to \mu\mu$ decays, with systematic uncertainties assigned to cover the extrapolation from \PZ to \PW boson decays. Ref.~\cite{Lupton} showed that variations in the event generator QCD tuning parameters \IKT and $\alpha_s$ have a contrasting effect on the muon \pt distribution, such that these parameters could also be floated in a simultaneous fit to \PW and \PZ boson data. In summary, we float $m_W$ and all the aforementioned nuisance parameters in a simultaneous fit of the muon $q/\pt$ distribution from $\PW \to \mu\nu$, and the  $\phi^{\ast}$~\cite{phistar} distribution from $Z \to \mu\mu$. The former is chosen over \pt for plotting convenience, and the latter over $\pt^Z$ for its insensitivity to detector modelling details. This fit model was validated by using our central model to fit \textit{pseudodata} generated with different models of the unpolarised cross section. The $m_W$ values found in these pseudodata fits had a similar spread to that found when using different models to fit the real data. This demonstrates that the fit model has sufficient flexibility to describe the underlying boson production and simultaneously extract $m_W$ with high precision. Finally, the measurement is performed with three recent global PDF sets: NNPDF3.1~\cite{Ball:2017nwa}, CT18~\cite{Hou:2019efy} and MSHT20~\cite{Bailey:2020ooq}. 

\section{Detector modelling and calibration}

Accurate preparation of the templates also requires that the detector response is well understood and well modelled. The first part of ensuring this is to correct for any misalignment of the detector at the analysis level, since this can bias our measurement of the muon \pt spectrum. After a custom detector alignment algorithm using high-\pt muons from \PZ boson decays is applied, we apply finer curvature ($q/p$) corrections derived by the \emph{pseudomass} method applied on $Z \to \mu\mu$ decays~\cite{Barter:2021npd}. 

\begin{figure}[!t!]\centering
  \includegraphics[width=\DefaultFigWidth]{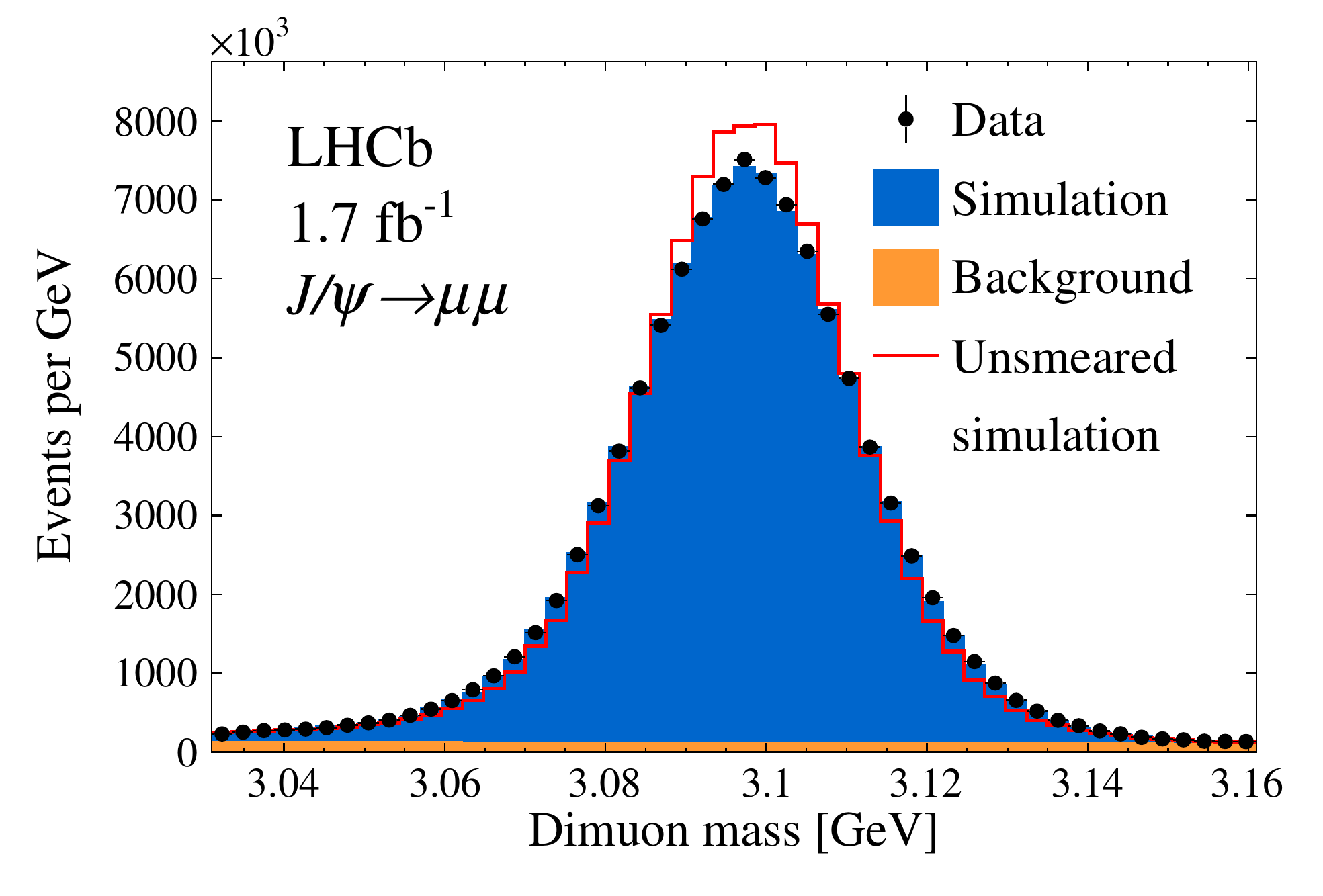}
  \includegraphics[width=\DefaultFigWidth]{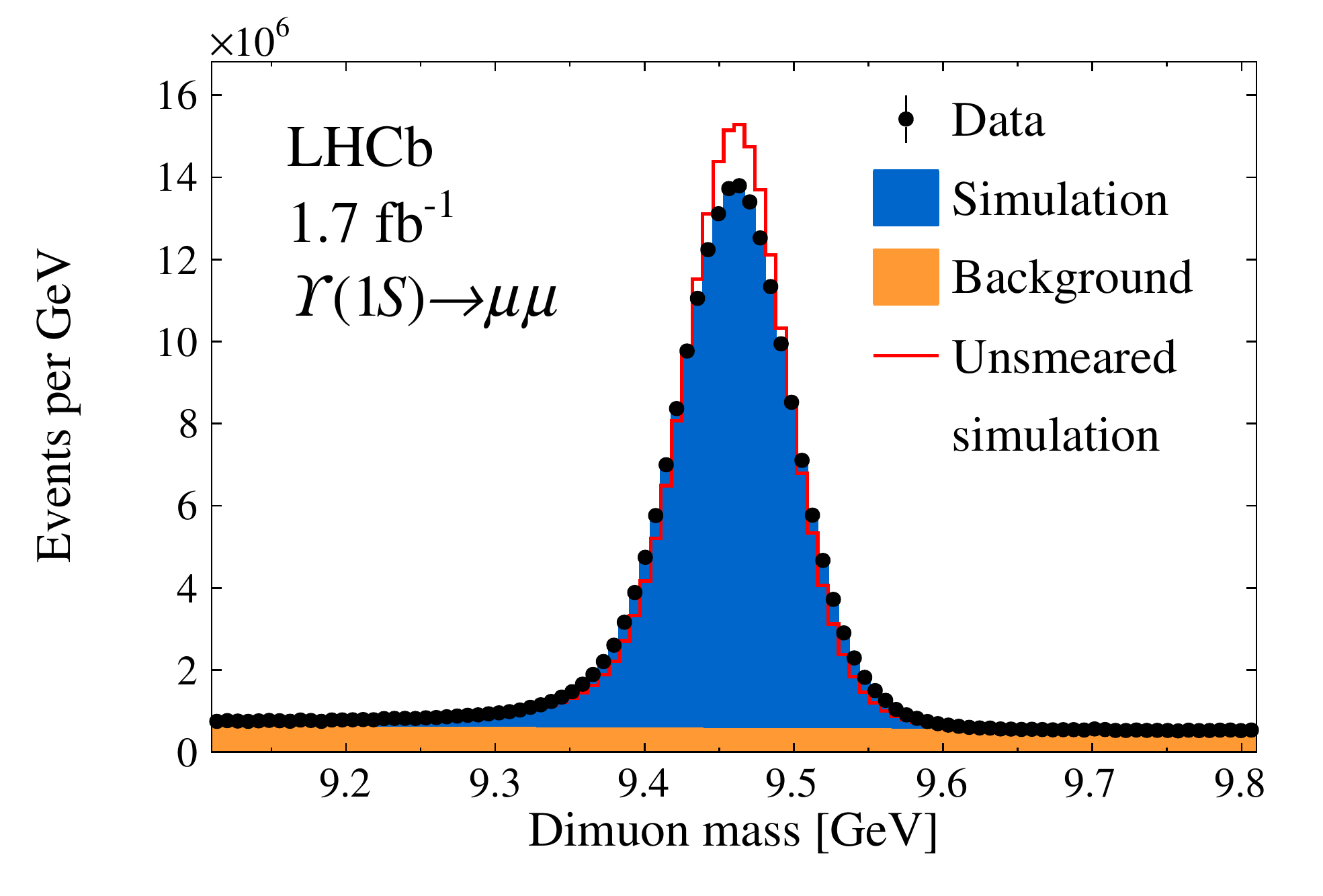}
  \includegraphics[width=\DefaultFigWidth]{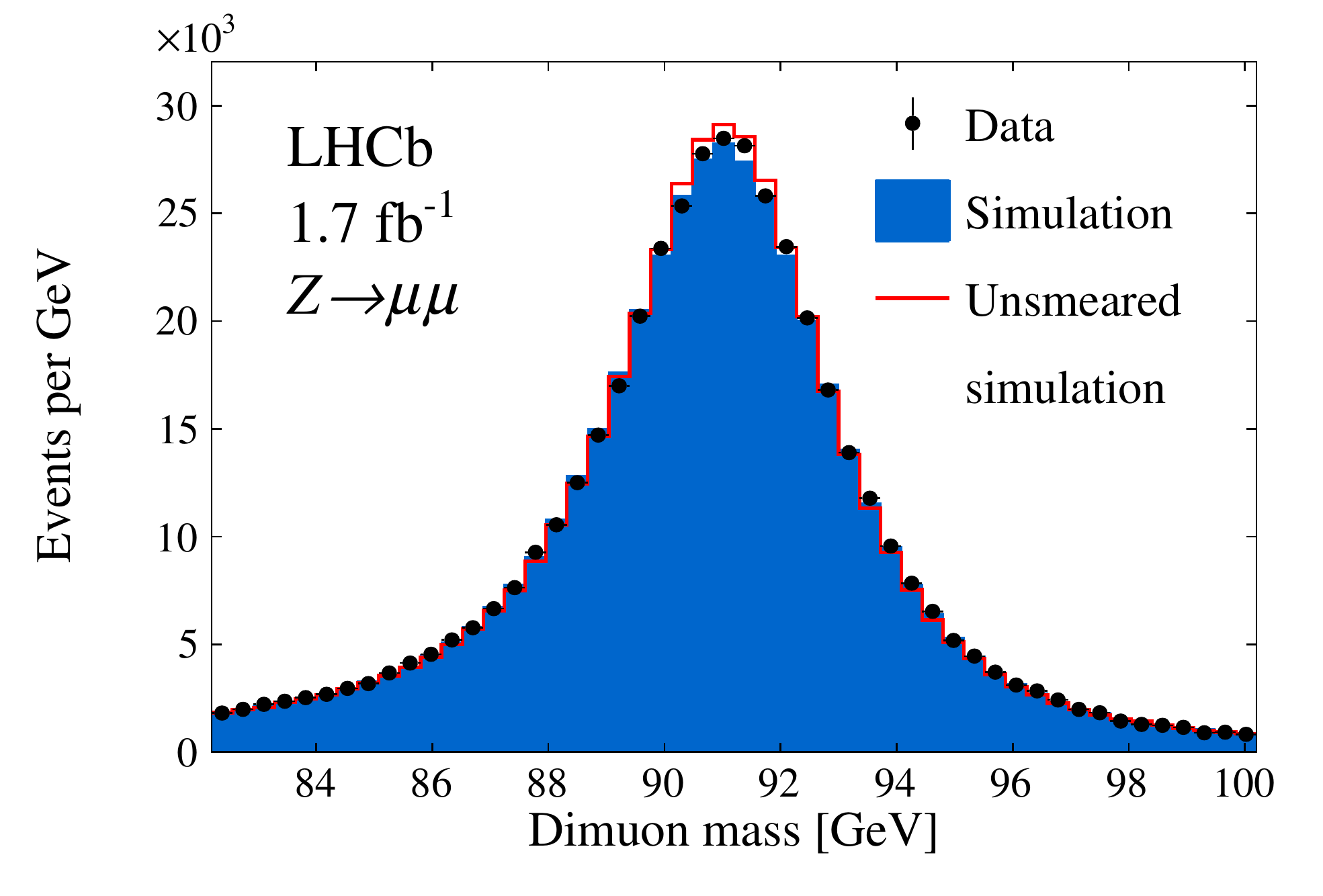}
  \caption{Fitted dimuon mass distributions for $J/\psi$, $\Upsilonres(1S)$ and \PZ boson candidates, combining all $\eta$ and magnet polarity categories, to determine the smearing parameters to be applied to the simulation. The red histogram indicates the model before the application of the smearing.}
\label{fig:MomentumScaleFit}
\end{figure}

These corrections are applied to both the data and simulation, giving the effect of realigning the detector for both. The simulated \lhcb momentum resolution and scale is then corrected with an additional smearing. Six smearing parameters are derived from 36 simultaneous fits (binned in magnet polarity and $\eta$) to the \jpsi, $\Upsilonres(1S)$ and \PZ boson invariant mass peaks. Re-combining these categories, the fit result is shown in Fig.~\ref{fig:MomentumScaleFit}.

A set of selection requirements are then applied to give a high purity sample of \PW and \PZ boson decays. The most important requirements in the \PW boson selection are a ``\PZ-veto'' on events where there is a second, high-\pt muon in the \lhcb acceptance; that there is a well-reconstructed track that is identified as a muon and fires high-\pt muon triggers; and that the muon is isolated from other particles in the event. Our fit range is $28 < \pt < 52$ GeV and $2.2 < \eta < 4.4$, which yields 2.4 million \PW boson candidates. Whilst the resulting \PZ boson sample is extremely pure, significant residual backgrounds are present in the \PW boson data. EW backgrounds such as $Z \to \tau\tau$ and $W \to \tau\nu$ can be fully simulated and constrained relative to the \PZ boson sample. The largest remaining background is that of light hadrons decaying in-flight to muons. This background is described with a parametric model that is trained on a hadron-enriched data sample. 

Each of the aforementioned selection requirements comes with an efficiency which may correlate with the muon \pt. If these efficiencies are mismodelled in the simulation, this leads to a bias on $m_W$. Mismodelling of the muon tracking, ID and trigger efficiencies are corrected for using the tag-and-probe method with dimuon \PZ boson and $\Upsilon(1S)$ control samples. Efficiencies are calculated (in data and simulation) as a function of muon \pt, $\eta$ and azimuthal angle $\phi$, and the templates are then corrected by weighting them according to the (binned) efficiency ratio $\varepsilon_{data}/\varepsilon_{sim}$. The muon isolation efficiency is handled in a similar way (with only \PZ boson decays): efficiencies (and hence corrections) are derived as a function of muon $\eta$ and \emph{recoil projection} $u = \vec{p}^{\;V}_{\rm T} \cdot \vec{p}^{\;\mu}_{\rm T} / p_{\rm T}^{\mu}$. 

\section{Fit Result and Uncertainties}

\begin{figure}\centering
\includegraphics[width=\DefaultFigWidth]{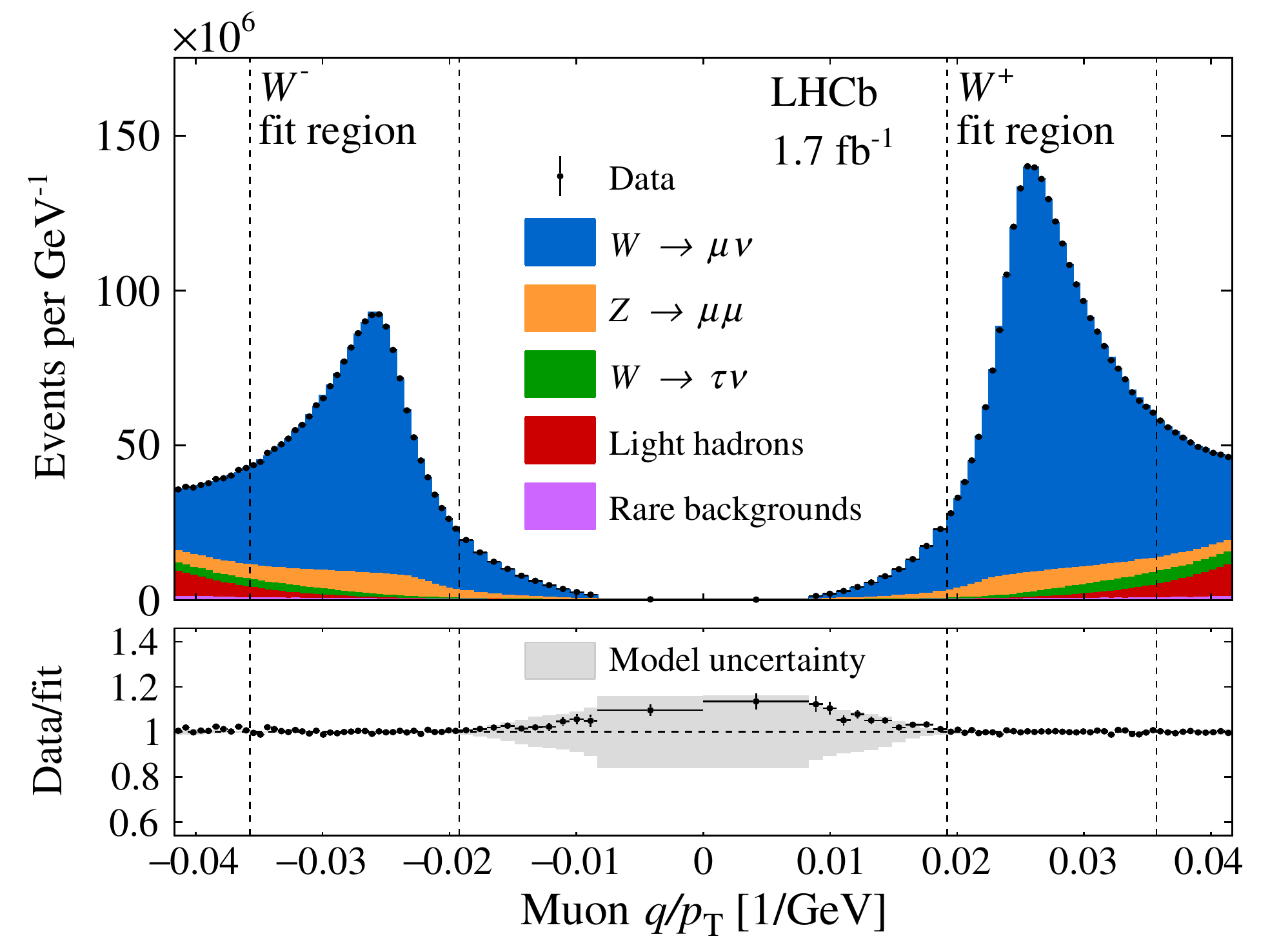}
\includegraphics[width=\DefaultFigWidth]{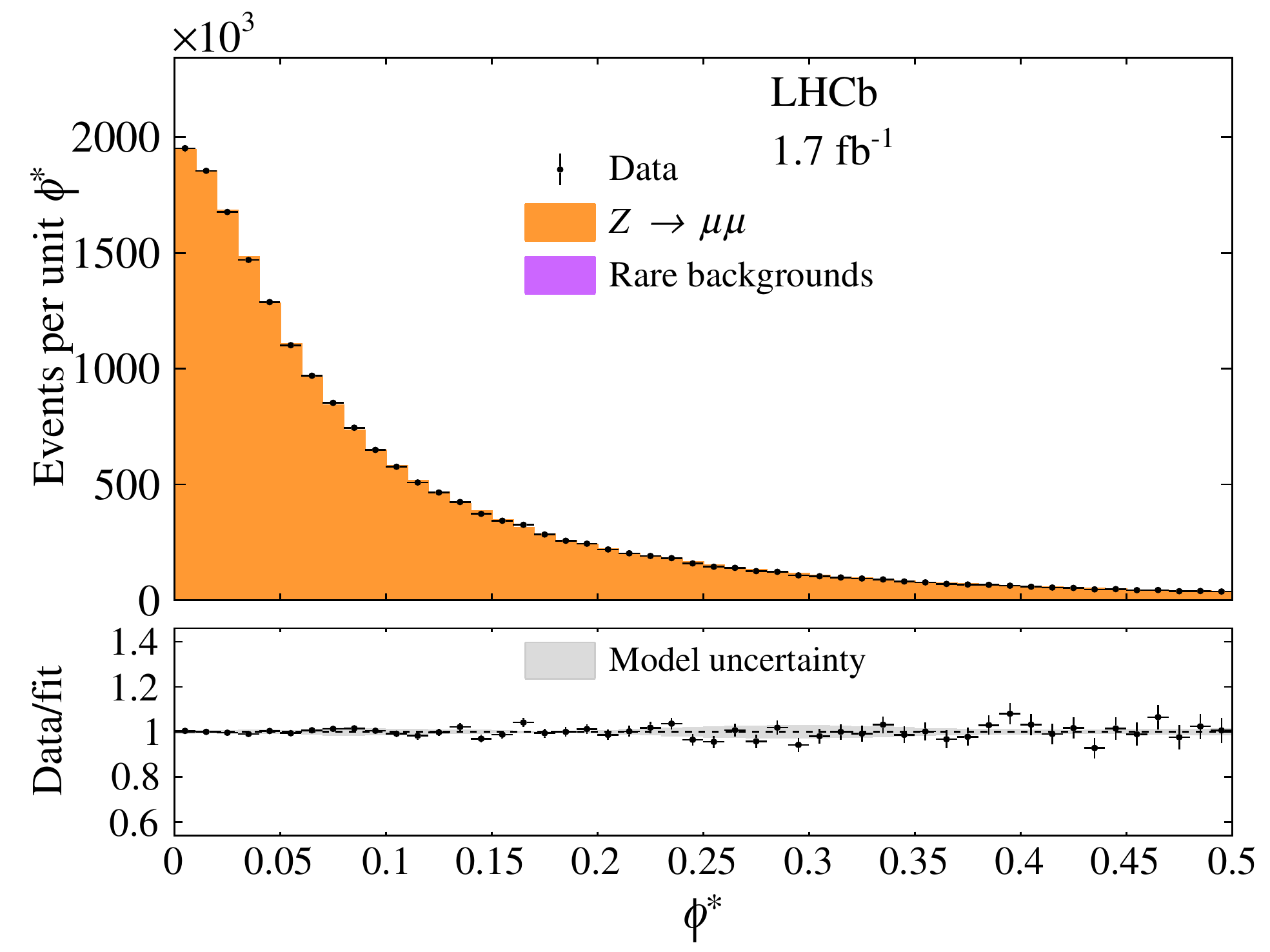}
\caption{\label{fig:remake}Distributions of (left) $q/\pt$ and (right) $\phi^{\ast}$ compared to the model after the $m_W$ fit.}
\end{figure}

Fig.~\ref{fig:remake} shows the data with the fitted model (NNPDF3.1 PDFs) overlaid. The statistical uncertainty on $m_W$ is 23\mev. The fit $\chi^2$ per degree of freedom is 105/102, and the $A_3$ scaling factor is consistent with unity at the $1\sigma$ uncertainty level.

\newcommand*{\tabindent}{ \hspace{3mm}}
\begin{table}\centering
\begin{tabular}{ll|ll}
Source & Size & Source & Size \\
\hline
Parton distribution functions & $\DataMWPDF$  & Experimental total & $\DataMWExp$ \\
Theory (excl. PDFs) total& $\DataMWTh$ & \tabindent Momentum scale and resolution & $7$ \\
\tabindent Transverse momentum model & $11$ & \tabindent Muon reco. efficiencies & $6$\\
\tabindent Angular coefficients & $10$ & \tabindent Isolation efficiency & $4$\\
\tabindent QED FSR model & $7$ & \tabindent QCD background & $2$\\
\tabindent Additional EW corrections & $5$ & & \\
\end{tabular}
\caption{\label{Tab:syst}Leading (above 1\mev) contributions to the systematic uncertainty in $m_W$ in\mev.}
\end{table}

\begin{figure}[ht]
    \centering
    \includegraphics[width=0.75\textwidth]{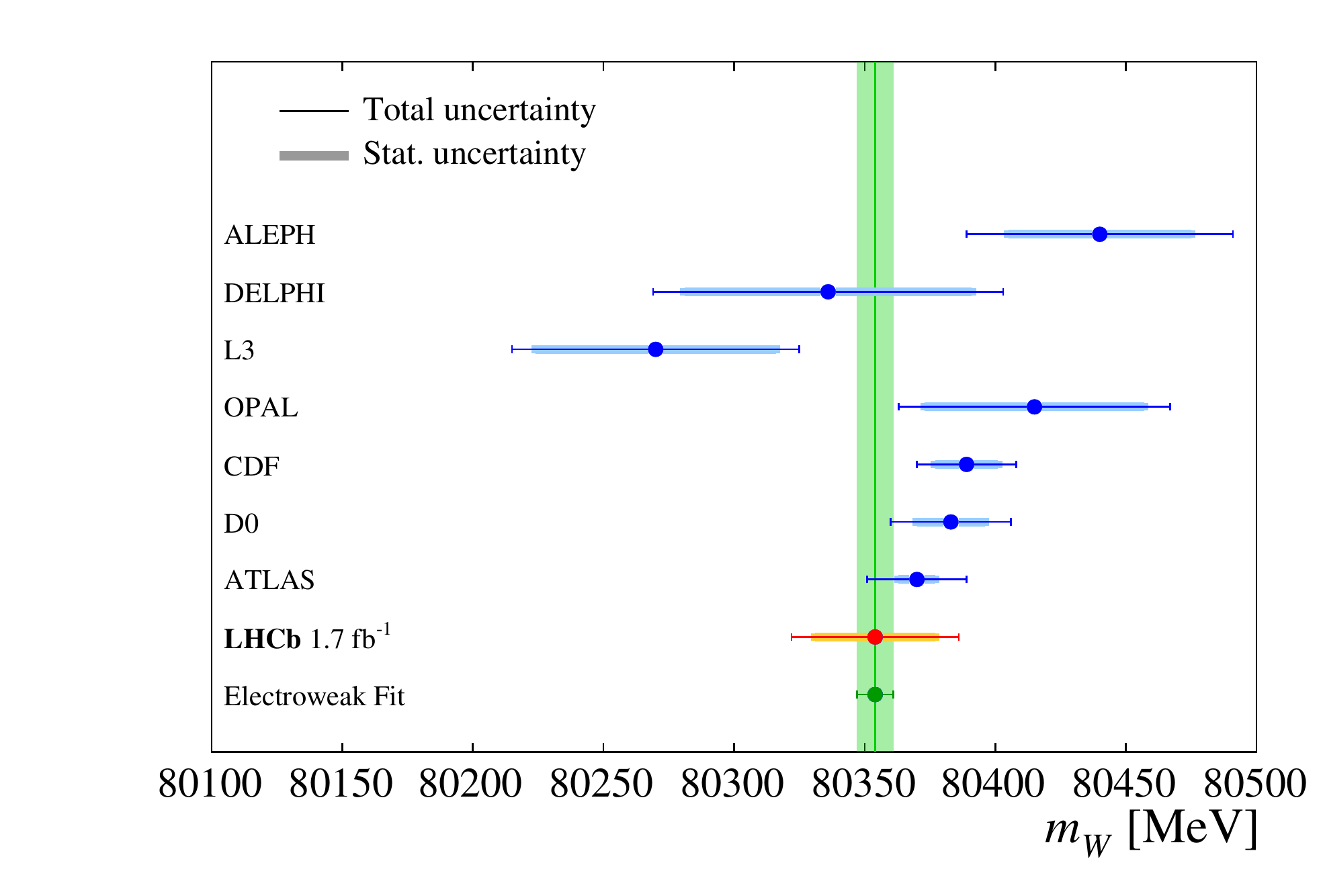}
    \caption{Measured value of $m_W$ compared to those from the \aleph~\cite{ALEPH:2006cdc}, \delphi~\cite{DELPHI:2008avl}, \lthree~\cite{L3:2005fft}, \opal~\cite{OPAL:2005rdt}, \cdf~\cite{CDF}, \dzero~\cite{D0} and \atlas~\cite{atlas} experiments. The current prediction of $m_W$ from the global electroweak fit~\cite{Gfitter} is also included.}
    \label{fig:mW_summary_experiments}
\end{figure}

A breakdown of the leading systematic uncertainties is given in Table~\ref{Tab:syst}. The boson \pt model and QED Final State Radiation (FSR) uncertainties are estimated by taking alternative predictions from different models/programs, whereas the angular coefficient uncertainty comes from uncorrelated scale variations, as recommended by Ref.~\cite{Gauld:2017tww}. An uncertainty of 5\mev is assigned for missing higher order EW corrections. The experimental systematic uncertainties are evaluated by varying modelling details (e.g. binning/smoothing of control samples, parametric shapes), propagating the statistical uncertainties from control samples, and uncertainties on external input values.
Since the three PDF sets use almost identical data, we make no preference between them and consider their uncertainties to be fully correlated. Therefore, measurements from each set are arithmetically averaged to produce our central result. The overall PDF uncertainty is also an arithmetic average of the three PDF uncertainties, evaluated according to the prescription of each PDF group. The resulting measurement is 

\begin{equation}
    m_W = \DataMWValue \pm \DataMWStat_{\textrm{stat}} \pm \DataMWExp_{\textrm{exp}} \pm \DataMWTh_{\textrm{theory}} \pm \DataMWPDF_{\textrm{PDF}} \,\mev,
\end{equation}
\noindent
with a total uncertainty of 32\mev. This result is compatible with the current PDG average of direct measurements~\cite{PDG2020} and the SM prediction from the global EW fit~\cite{Gfitter}. It is compared to previous measurements in Fig.~\ref{fig:mW_summary_experiments}. 

\section{Conclusion and Outlook}

In these proceedings, a first measurement of the \PW boson mass by \lhcb has been presented~\cite{WmassPaper}. A total uncertainty of approximately 32\mev is achieved, despite using roughly one third of the \lhcb Run-2 dataset. This proof-of-principle measurement shows that a ${\sim}20\mev$ total uncertainty is achievable using the full dataset. Ref.~\cite{Pili} has already shown that the PDF uncertainty can be substantially reduced using \textit{in situ} PDF constraints and by fitting the doubly differential \pt and $\eta$ distributions, rather than just \pt. However, particular effort is needed to reduce the dominating systematic uncertainties in the modelling of the boson production and decay. 

\bibliographystyle{JHEP}
\bibliography{main}

\end{document}